\documentclass[
]{ceurart}

\usepackage{listings}
\usepackage{hyperref}
\usepackage{graphicx}
\usepackage{todonotes}
\usepackage{subcaption}
\usepackage{soul}

\usepackage{tikz}
\usepackage{tikz-qtree}
\usetikzlibrary{trees} 

\newcommand\fs[1]{\href{https://fairsharing.org/#1}{[fs:#1]}}
\newcommand\rd[1]{\href{https://www.re3data.org/repository/#1}{[rd:#1]}}
\newcommand\od[1]{\href{https://v2.sherpa.ac.uk/id/repository/#1}{[od:#1]}}
\newcommand\rr[1]{\href{http://roar.eprints.org/#1}{[rr:#1]}}

\begin{document}

\copyrightyear{2023}
\copyrightclause{Copyright for this paper by its authors.
  Use permitted under Creative Commons License Attribution 4.0
  International (CC BY 4.0).}

\conference{IRCDL'23: The Conference on Information and Research science Connecting to Digital and Library science 2023 (Formerly the Italian Research Conference on Digital Libraries) 23-24 February 2023 - Bari, Italy}

\title{(Semi)automated disambiguation of scholarly repositories}
\tnotemark[1]
\tnotetext[1]{All authors contributed equally.}

\author[1]{Miriam Baglioni}[%
orcid=0000-0002-2273-9004,
email=miriam.baglioni@isti.cnr.it
]

\author[1]{Andrea Mannocci}[%
orcid=0000-0002-5193-7851,
email=andrea.mannocci@isti.cnr.it
]
\fnmark[1]

\author[1]{Gina Pavone}[%
orcid=0000-0003-0087-2151,
email=gina.pavone@isti.cnr.it
]

\author[1]{Michele De Bonis}[%
orcid=0000-0003-2347-6012,
email=michele.debonis@isti.cnr.it
]

\author[1,2]{Paolo Manghi}[%
orcid=0000-0001-7291-3210,
email=paolo.manghi@isti.cnr.it
]

\address[1]{CNR-ISTI -- National Research Council, Institute of Information Science and Technologies ``Alessandro Faedo'', 56124 Pisa, Italy}
\address[2]{OpenAIRE AMKE, Athens, Greece}

\fntext[1]{Corresponding author.}

\begin{abstract}
The full exploitation of scholarly repositories is pivotal in modern Open Science, and scholarly repository registries are kingpins in enabling researchers and research infrastructures to list and search for suitable repositories.
However, since multiple registries exist, repository managers are keen on registering multiple times the repositories they manage to maximise their traction and visibility across different research communities, disciplines, and applications.
These multiple registrations ultimately lead to information fragmentation and redundancy on the one hand and, on the other, force registries' users to juggle multiple registries, profiles and identifiers describing the same repository.
Such problems are known to registries, which claim equivalence between repository profiles whenever possible by cross-referencing their identifiers across different registries.
However, as we will see, this ``claim set'' is far from complete and, therefore, many replicas slip under the radar, possibly creating problems downstream.
In this work, we combine such claims to create duplicate sets and extend them with the results of an automated clustering algorithm run over repository metadata descriptions. Then we manually validate our results to produce an ``as accurate as possible'' de-duplicated dataset of scholarly repositories.
\end{abstract}

\begin{keywords}
  Scholarly Registries \sep Scholarly Repositories \sep De-duplication \sep Open Science
\end{keywords}

\maketitle
\section{Introduction}
Scholarly repositories are essential for Open Science practice, as they enable access to research products and grant their long-term preservation. Besides, they play a crucial role in improving the visibility, discoverability and reuse of research products~\cite{pampel2013,wallis2013,davidson2014,pasquetto2019,silvello2018}.

Given the ever-increasing number of repositories over the years, there is a growing need for \textit{scholarly repository registries} providing repositories with an \textit{identity} to enable non-ambiguous reference, support provenance tracking and impact assessment.
Moreover, registries hold authoritative descriptive \textit{profiles} of repositories intended to foster their discoverability.

To this end, specialised scholarly repository registries have been set up~\cite{ball2014,pampel2013,wallis2013,davidson2014} in order to store a broad range of information about registered repositories, such as the type of content, discipline and subjects, access rights, and licensing information of the resources they store.

As a plurality of registries does exist and serves different scientific domains, communities and use cases, repository managers are incentivised to register their repository in more than one registry to boost their online presence and traction at the expense of information maintainability and up-to-dateness.
Such a variety of scholarly repository registries provides a full-spectrum overview across scientific disciplines and research applications and becomes a rich asset for scholarly registries' users downstream, such as researchers, scholarly service providers, and Open Science infrastructures listing and aggregating their content. 

However, the availability of multiple registries, repository identifiers and profiles poses non-trivial questions and challenges regarding their authoritativeness, disambiguation, and coverage.
In particular, such fragmentation quickly becomes a drawback in terms of information redundancy and scattering, as multiple registrations produce different identifiers for the same repository, arbitrarily used across the scholarly communication track record, and may lead to potential information inconsistencies across registries. 

Unsurprisingly, registry managers are aware of such drawbacks and claim equivalence of repository profiles whenever possible by cross-referencing their identifiers and PIDs across different registries.
As we will see, however, this ``claim set'' is far from complete. 
Nevertheless, this is the only ``ground knowledge'' at our disposal; hence, we consider it valid without further challenge.

In this work, we conflate such claims to infer the duplicate sets of different profiles about the same repository across the registries.
Then, we further extend such duplicate sets by integrating them with the clusters obtained by an automated clustering algorithm run over repository profiles.
Finally, we deliver the dataset of repository duplicates, taking care of manually validating all the cases where the de-duplication contributed to integrating the claims provided by the registries, and we draw some conclusions.

\section{Related work}
\label{sec:soa}
The many scholarly communication services and academic search systems spun in the last decade have grown in parallel with a different, often siloed, mindset to target a broad and diverse range of use cases and application contexts~\cite{martin-martin2020,gusenbauer2020,visser2021,harzing2019,aryani2020}.
Consequently, despite modelling very similar aspects of the academic domain (often the same ones), the respective data models ended up being quite distant in order to capture the different peculiarities at hand. 
Scholarly registries are no exception as they have often been developed to target diverse research communities, academic disciplines and research applications, as can be easily derived by inspecting the documented data models and schemas, as documented in Section~\ref{sec:datamethods}.

Intuitively, talking about their content comparison and interconnection is paramount to ensure their consistency and pave the way towards full interoperability and information exchange across scholarly registries.

To the best of our knowledge, no prior study analysed and compared the content of publicly available scholarly registries and highlighted their inherent ambiguity.
Therefore, in this work, we address the disambiguation of scholarly repositories across scholarly repository registries, which are at the centre of the present study.

To date, the only tangible trace towards this direction can be found in the claims provided in some cases by scholarly registries to cross-reference other repository registration across other registries.
As we will see further into our analysis, these efforts are, however, not enough, and the ``claim set'' provided by the registries is far from ideal, as many subsequent registrations of the same research repositories went so far unnoticed.

\section{Data and methods}
\label{sec:datamethods}
For this study, we selected four prominent scholarly repository registries, namely FAIRsharing\footnote{FAIRsharing -- \url{https://fairsharing.org}}~\cite{sansone2019}, re3data\footnote{re3data registry -- \url{https://re3data.org}}~\cite{pampel2013}, OpenDOAR\footnote{OpenDOAR registry -- \url{https://v2.sherpa.ac.uk/opendoar}}, and ROAR\footnote{ROAR registry -- \url{http://roar.eprints.org/information.html}}, whose details are summarised in Table~\ref{tab:data} and the following.

\paragraph{FAIRsharing}
\label{sec:fairsharing}
Hosted at the University of Oxford in the UK and launched in 2011, FAIRsharing~\cite{sansone2019} is a web-based, searchable portal of three interlinked registries, containing both in-house and crowdsourced, manually curated descriptions of standards, databases (including repositories and knowledge bases) and data policies, which are persistently identifiable via DOIs.
FAIRsharing maps the landscape of these three resources, monitoring their relationships, development, evolution and integration, such as the implementation and use of standards in databases or their adoption in data policies by funders, journals and other organisations. 
FAIRsharing is also an endorsed output of the RDA FAIRsharing WG\footnote{RDA FAIRsharing WG -- \url{https://www.rd-alliance.org/group/fairsharing-registry-connecting-data-policies-standards-databases.html}}, and its management combines a community-driven approach where the internal curators are supported by the maintainers of the resources themselves, which get credited via their ORCID. 

As of February 2022, FAIRsharing has over 3,600 resources, of which 1,853 are databases, i.e., scholarly repositories of interest for this analysis.
The content, licenced under a CC-BY-SA licence\footnote{FAIRsharing licence -- \url{https://fairsharing.org/licence}}, can be browsed by (among other fields) registry type, discipline and country, and a live statistics page provides several at-glance-views of the landscape\footnote{FAIRsharing stats -- \url{https://fairsharing.org/summary-statistics}}.

\paragraph{re3data}
\label{sec:re3data}
re3data\footnote{re3data registry -- \url{https://re3data.org}}~\cite{pampel2013} is a global registry of research data repositories from all academic disciplines. 
Since its launch in 2012, this registry has been funded by the German Research Foundation\footnote{German Research Foundation (DFG) -- \url{https://dfg.de}} (DFG). 
The urgency of avoiding duplication of effort and serving the research community with a single, sustainable registry is mentioned on the ``About'' web page of the website registry, referring to the merger between re3data and Databib in 2013. 
Under the project ``re3data.org -- Community Driven Open Reference for Research Data Repositories (COREF)'', the registry provides ``customisable and extendable core repository descriptions that are persistently identifiable and can be referred to and cited appropriately. This includes unique identification in machine-to-machine communication''\footnote{re3data mission statement -- \url{https://www.re3data.org/about}}.
re3data was created to meet the need for a resource specifically dedicated to data repositories~\cite{pampel2013}.
In fact, the already existing registries OpenDOAR and ROAR mainly focused on repositories for scholarly publications and only hosted a residual share of data repositories. Furthermore, there was a need for a more detailed description of each research data repository, e.g., containing precise information on access and reuse conditions.
The registry can be browsed by content type, discipline or country.

For convenience, we retrieved the data via the OpenAIRE project\footnote{OpenAIRE -- \url{https://www.openaire.eu}} as it already integrates the registry.
As of February 2022, it contains 2,793 data repository profiles; the registry content is released under a CC-BY licence.
The collected data follow the re3data schema version 2.2\footnote{re3data.org 2.2 schema -- \url{https://www.re3data.org/schema/2-2}}.
On August 2021, re3data delivered a new schema version\footnote{re3data 3.1 schema -- \url{https://www.re3data.org/schema}}, and we used this one for the crosswalk.

\paragraph{OpenDOAR}
\label{sec:opendoar}
OpenDOAR\footnote{OpenDOAR registry -- \url{https://v2.sherpa.ac.uk/opendoar}} is a directory listing only Open Access repositories. The service was launched in 2005 as a result of a collaboration between the University of Nottingham and Lund University, funded by OSI, Jisc, SPARC Europe and CURL.
The listed repositories are grouped into five types (Undetermined, Institutional, Disciplinary, Aggregating, Governmental), which can be browsed by type of content, software, countries and regions\footnote{OpenDOAR advanced browsing -- \url{https://v2.sherpa.ac.uk/cgi/search/repository/advanced}}.
To be included in OpenDOAR, repositories must meet the inclusion criteria, and the submission has to be accepted by the curators' team. The submission request to OpenDOAR is carried out in two parts: first, the application is sent by filling in a form with basic information. If the admission criteria are met, further information is requested.

We retrieved the data via OpenAIRE as it natively integrates the registry content. 
As of Feb 2022, OpenDOAR lists 5,811 repositories worldwide; the content of the registry is redistributed under a CC-BY-NC-ND licence.
The collected data follow the schema accessible through their website\footnote{OpenDOAR schema --  \url{https://v2.sherpa.ac.uk/api/metadata-schema.html}}.

\paragraph{ROAR}
\label{sec:roar}
The Registry of Open Access Repositories (ROAR) is hosted at the University of Southampton, UK, and it is funded by the Jisc\footnote{ROAR registry -- \url{http://roar.eprints.org/information.html}}.
As declared on the project's web page, its aim is ``to promote the development of Open Access by providing timely information about the growth and status of repositories throughout the world''.
A registered account is needed to add a new repository, and the submission will be reviewed and eventually accepted or rejected with editorial comments\footnote{ROAR registration -- \url{http://roar.eprints.org/cgi/roar\_register}}. 

The data have been downloaded directly from the site, choosing the option \textit{Multiline CSV}. 
The CSV data formatting is consistent with the schema available online\footnote{ROAR schema -- \url{http://roar.eprints.org/cgi/schema}}.
As of February 2022, it contains 5,444 data repository profiles.
The registry's content can be browsed by country, year, repository type, institutional association, repository software\footnote{ROAR stats -- \url{http://roar.eprints.org/view}} is redistributed under a CC-BY licence.

\paragraph{}
As mentioned, in some cases, three out of four registries provide \textit{claims} (sometimes provided by users), establishing ``same-as'' equivalences among repository profiles registered across registries.
However, the claim set is far from complete, as FAIRsharing provides claims to re3data only, ROAR provides claims to OpenDOAR only, and re3data provides claims to all the other three (often not at the same time, though).
Moreover, we empirically noticed that no registry provides claims addressing internal duplicates (see Section~\ref{sec:results} for more details). 
Finally, no assurance about their correctness nor the existence of missing claims is given.
Yet, such claims are the only ``ground knowledge'' at our disposal; hence, we consider them valid without further challenge.
\begin{table*}
  \caption{Summary of the four registries}
  \label{tab:data}
  \begin{tabular}{rlllll}
    \toprule
    Registry    & Dump date & Dump method     & Licence     & \# repos & \# claims \\
    \midrule
FAIRsharing & Feb 2022  & JSON (rest API) & CC-BY-SA    & 1,853    & 787       \\
re3data     & Feb 2022  & OpenAIRE        & CC-BY       & 2,793    & 475       \\
OpenDOAR    & Feb 2022  & OpenAIRE        & CC-BY-NC-ND & 5,811    & N/A       \\
ROAR        & Feb 2022  & CSV (website)   & CC-BY       & 5,444    & 3,279    \\
  \bottomrule
\end{tabular}
\end{table*}

In order to address such shortcomings and identify further repository duplicates, in the present work, we first use the claims provided by the three registries and conflate them to create duplicate sets whenever they regard different profiles of the same (allegedly) repository.
Then, we further extend the duplicate sets by integrating them with the clusters obtained by an automated clustering algorithm run over repository profiles.
Whenever a duplicate set intersects with one or more clusters, we try to extend it with new repository profiles provided by the clusters; otherwise, it is left untouched. 
Finally, all the clusters not intersecting with any duplicate set are promoted to duplicate sets.
As the last step, a manual validation is performed for all the duplicate sets where the de-duplication is involved.
All the duplicate sets obtained as such, together with the code and original data, can be found in our Gitea repository\footnote{Data and code -- \url{https://code-repo.d4science.org/miriam.baglioni/Registries}}.

\begin{figure}[t]
    \centering
    \includegraphics[width=\linewidth]{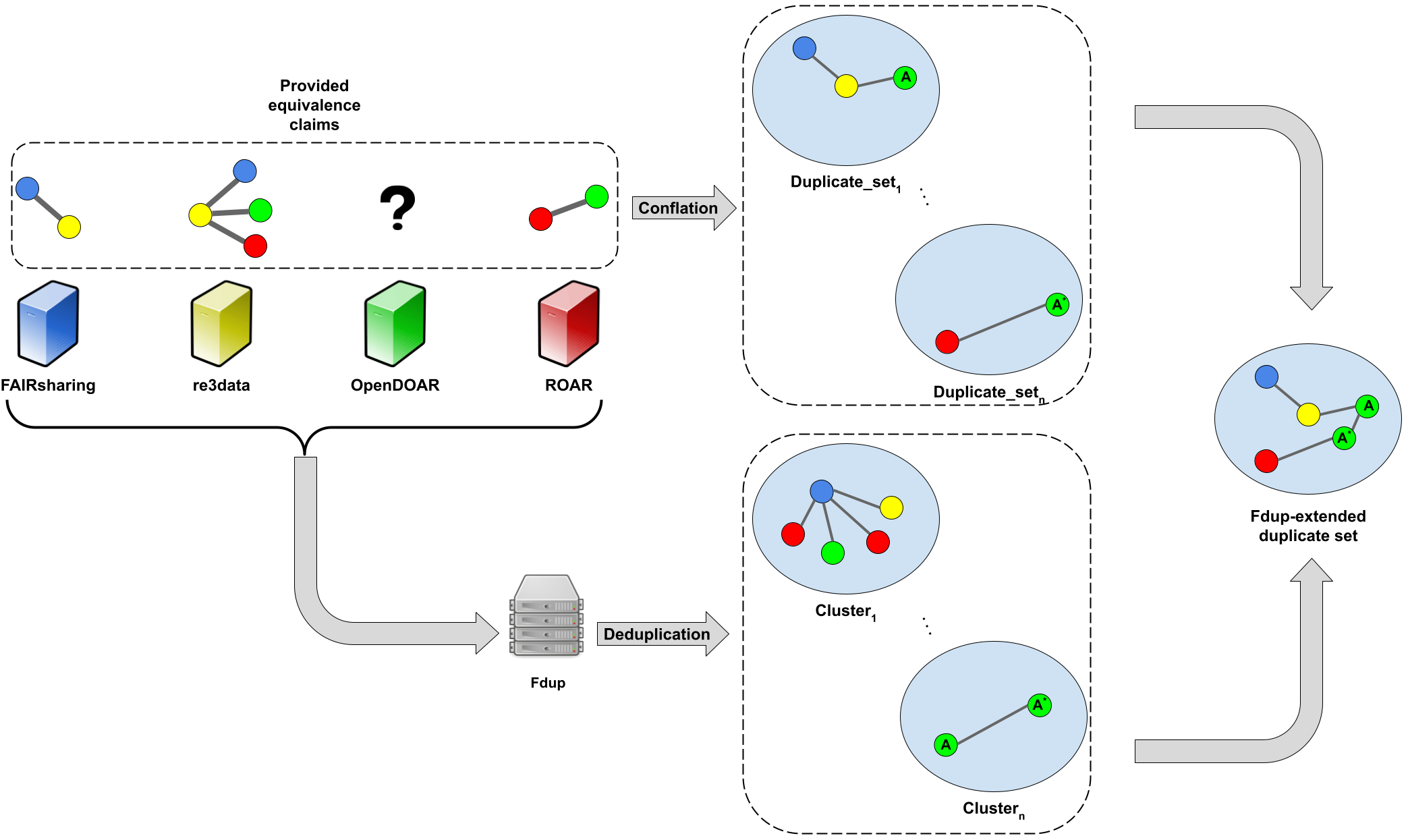}
    \caption{Overview of the methodology. Firstly, the claims provided by the registries are conflated in order to derive duplicate sets; then, the identified duplicate sets are extended with the clusters provided by the FDup automatic de-duplication algorithm running against the content of the registries. In this example, FDup discovers an equivalence between repository profiles $A$ and $A^*$, and this triggers a merger between the two duplicate sets represented above.}
    \label{fig:methods}
\end{figure}

For the sake of clarity, the methodology is represented in Figure~\ref{fig:methods} and described in details in the following paragraphs.

\paragraph{Conflating registry claims}
To conflate the claims, we arbitrarily start from FAIRsharing and programmatically explore, for each repository profile in FAIRsharing, the claims referring to repository profiles in re3data (e.g., \fs{2114}$\rightarrow$\rd{r3d100010191})\footnote{Hereafter, we refer to repository profiles by indicating repository registration identifiers prefixed according to the involved registry (i.e., fs: -- od: -- rd: -- rr:). Each example links to the profile in the relevant registry in its current version, which might differ from the one observed at the time of writing. The metadata as we collect them are provided for transparency and validation of the reported examples.} to seed the starting duplicate sets.
Then, for any given duplicate set, we verify whether it exists a claim from the re3data profile pointing back to the starting FAIRsharing repository (e.g., \fs{2114}$\rightarrow$\rd{r3d100010191}$\rightarrow$\fs{2114}). 
If this is not the case, we note down the three profiles involved as a \textit{problematic} duplicate set, which has to be controlled manually at a later stage (e.g., \fs{3652}$\rightarrow$\rd{r3d100012729}$\rightarrow$ \fs{1724}), and we move forward.

Next, we try to extend any given duplicate set by conflating the re3data profile claims towards OpenDOAR. 
A new profile from OpenDOAR is added to a duplicate set if and only if it is not already part of another; otherwise, the duplicate set is marked as problematic and will be controlled manually. 
Then the claims pointing from the re3data profile to ROAR are conflated, and, as for OpenDOAR, one profile from ROAR is added to the duplicate set only if it is not already part of another set; otherwise, the set will be controlled manually.

If one profile from ROAR is added to a duplicate set, we consider its claims. Since ROAR provides claims only towards OpenDOAR, a profile from OpenDOAR could be added to the duplicate set if it is not already present in another set.

Finally, for each re3data claim not already processed, we seed a new duplicate set and search for FAIRsharing, OpenDOAR, and ROAR claims as done before.
Similarly, we check the ROAR repositories claims towards OpenDOAR that have not already been processed. 
In this case, if the OpenDOAR profile is present in another duplicate set, we extend the already formed duplicate set with the new ROAR profile\footnote{Multiple profiles from ROAR can be added to the same duplicate set because more than one ROAR repository can claim the same OpenDOAR one, e.g., \rr{919} and \rr{5425} both claim \od{1047}}.

\paragraph{De-duplication}
In order to automatically detect repository duplicates and cluster them, we opted for FDup~\cite{de2022fdup}.
FDup has been developed within the effort of the OpenAIRE-Advance\footnote{OpenAIRE-Advance -- \url{https://cordis.europa.eu/project/id/777541}} and OpenAIRE-Nexus\footnote{OpenAIRE-Nexus -- \url{https://cordis.europa.eu/project/id/101017452}} projects and is currently applied in the production of the OpenAIRE Research Graph~\cite{manghi2021} to de-duplicate metadata records of research products and organisations.

FDup provides an efficient de-duplication framework capable of comparing the metadata records in a ``big data'' collection to identify the groups of equivalent ones, hereafter referred to as \textit{duplicate sets}.

To this aim, FDup is given a dataset consisting of all the repository profiles across the four registries, for which we selected a handful of relevant, common fields: original identifier and source registry for identification, repository name, and homepage URL. 
Then, FDup performs a two-phase processing: the \textit{candidate identification phase} optimises the otherwise quadratic complexity of comparing all possible pairs of profiles by pre-fetching the pairs of profiles that are likely representing the same repository; the \textit{duplicate sets identification} matches all pairs of candidate profiles to effectively validate their equivalence and finally generates the duplicates sets by grouping all identified matching pairs of profiles via their transitive closure.

The outcome of this process is a graph, where two nodes (i.e., repository profiles) are connected by an edge whenever similar.
The graph is then explored to identify its \textit{connected components}, which are the subgraphs in which each pair of nodes is connected via a path of edges (i.e., duplicate sets). 
The outcome of this phase strongly depends on the threshold chosen for the similarity match. 
For our analysis, we chose a 0.9 threshold to increase the precision level of the duplicate sets containing equivalent repositories.



\section{Result}
\label{sec:results}
While ROAR and FAIRsharing provide just one-to-one claims, re3data can claim duplicates in more than one registry. From re3data, in fact, we get one set composed of all four registries (e.g., \rd{r3d100012322}) and four sets composed of three registries out of four (e.g., \rd{r3d100012274}, \rd{r3d100013438}, \rd{r3d100013665}, and \rd{r3d100013717}). The remaining claims are one-to-one: 451 claims to FAIRsharing, 15 to OpenDOAR, and 4 to ROAR. 

\begin{figure}[t]
    \centering
    \begin{tikzpicture}[grow'=right,level distance=1.7in,sibling distance=.15in]
    \tikzset{edge from parent/.style={thick, draw, edge from parent fork right},
             every tree node/.style={draw,minimum width=1in,text width=1.3in,align=center,rounded corners=5}}
    \Tree 
        [.{3,548 \textbf{(100\%)}\\duplicate sets} 
            [.{3,151 \textbf{(88.8\%)}\\$N=2$}
                [.{2,337 \textbf{(74\%)}\\ROAR + OpenDOAR} ]
                [.{802 \textbf{(25\%)}\\FAIRsharing + re3data} ]
                [.{12 \textbf{(1\%)}\\re3data + \\OpenDOAR or ROAR } ]
            ]
            [.{348 \textbf{(9.8\%)}\\$N=3$}
                [.{341 \textbf{(98\%)}\\1x OpenDOAR + 2x ROAR} ]
                [.{7 \textbf{(2\%)}\\others} ]
            ] 
            [.{49 \textbf{(1.4\%)}\\$4 \leq N \leq 5$} ]
        ]
    \end{tikzpicture}
    \caption{Composition of the duplicate sets after conflating registries' claims. $N$ indicates the number of repository profiles participating in the same duplicate set.}
    \label{fig:results_claims}
\end{figure}
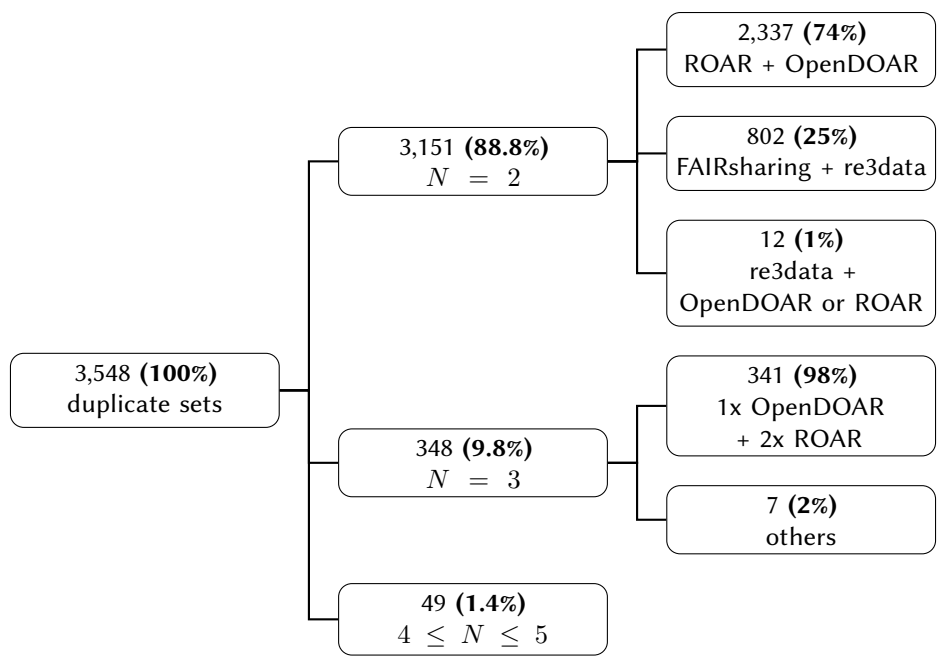

After conflating the registries' claims, we get 3,548 duplicate sets, whose composition is reported in Figure~\ref{fig:results_claims}. 
The 88.8\% of these sets consist of two profiles. As expected, the majority of the duplicate sets are composed of coupled profiles from OpenDOAR and ROAR (74\%), followed by FAIRsharing and re3data (25\%). The remaining involve duplicate sets where a re3data profile is paired with one in either ROAR or OpenDOAR. 

Sets composed of three profiles are 9.8\% of the total, and the majority (98\%) involve one profile in OpenDOAR and two in ROAR, suggesting the presence of duplicate repository registrations within ROAR. The remaining 2\% comprehends profiles drawn from re3data, OpenDOAR, and ROAR.

Finally, the last 1.4\% are sets composed of four or five profiles. In all but one, ROAR profiles appear more than once, up to four times in one set with OpenDOAR. Also, one duplicate from OpenDOAR is present. Summing up, 13.45\% of the repositories registered in ROAR, which provide a claim, can be considered duplicates (8.1\% of the total number of repositories).

The conflation produced six problematic duplicate sets, which involve only claims between FAIRsharing and re3data:
\begin{itemize}
    \item \fs{3652}$\rightarrow$\rd{r3d100012729}$\rightarrow$\fs{1724}. \fs{1724} was deprecated and has been subsumed into \fs{3652}. The set is \fs{3652} and \rd{r3d100012729}.
    \item \fs{3340}$\rightarrow$\rd{r3d100010543}$\rightarrow$\fs{2107}. \fs{2117} was deprecated and has been replaced by \fs{3340}. The set is \fs{3340} and \rd{r3d100010543}.
    \item \rd{r3d100010412}$\rightarrow$\fs{2424}$\rightarrow$\rd{r3d100011538}. \rd{r3d100011538} is a duplicate of \rd{r3d100010412}; the set can be considered right 
    \item \rd{r3d100011257}$\rightarrow$\fs{1730}$\rightarrow$\rd{r3d100012862}. \rd{r3d100011257} has been merged into \rd{r3d100012862}; the set is \rd{r3d100012862} and \fs{1730}
    \item \rd{r3d100011343}$\rightarrow$\fs{2163}$\rightarrow$\rd{r3d100000039}. \rd{r3d100011343} do not belong to the same set, and it refers to a different organisation. The set is: \fs{2163} and \rd{r3d100000039}
    \item \rd{r3d100013223}$\rightarrow$\fs{2524}$\rightarrow$\rd{r3d100012397}. \rd{r3d100013223} is a duplicate of \rd{r3d100012397}; the set can be considered to be right
\end{itemize}
After this manual validation, the checked claims are then used to extend the duplicate sets.

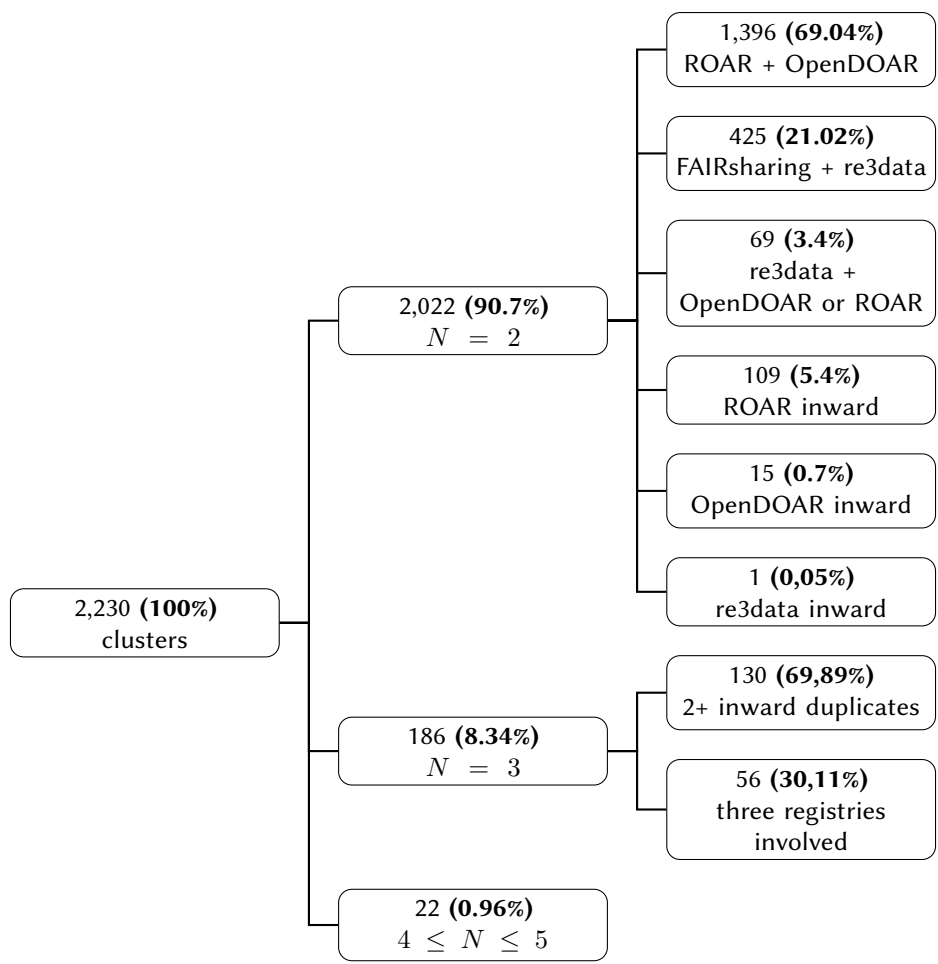
\begin{figure}[t]
    \centering
    \begin{tikzpicture}[grow'=right,level distance=1.7in,sibling distance=.15in]
    \tikzset{edge from parent/.style={thick, draw, edge from parent fork right},
             every tree node/.style={draw,minimum width=1in,text width=1.3in,align=center,rounded corners=5}}
    \Tree 
        [.{2,230\textbf{ (100\%)}\\clusters} 
            [.{2,022 \textbf{(90.7\%)}\\$N=2$}
                [.{1,396 \textbf{(69.04\%)}\\ROAR + OpenDOAR} ]
                [.{425 \textbf{(21.02\%)}\\FAIRsharing + re3data} ]
                [.{69 \textbf{(3.4\%)}\\re3data + \\OpenDOAR or ROAR } ]
                [.{109 \textbf{(5.4\%)}\\ROAR inward} ]
                [.{15 \textbf{(0.7\%)}\\OpenDOAR inward} ]
                [.{1 \textbf{(0,05\%)}\\re3data inward} ]
            ]
            [.{186 \textbf{(8.34\%)}\\$N=3$}
                [.{130 \textbf{(69,89\%)}\\2+ inward duplicates} ]
                [.{56 \textbf{(30,11\%)}\\three registries\\involved} ]
            ] 
            [.{22 \textbf{(0.96\%)}\\$4 \leq N \leq 5$} ]
        ]
    \end{tikzpicture}
    \caption{Composition of the duplicate clusters automatically identified by FDup. $N$ indicates the number of repository profiles participating in the same cluster.}
    \label{fig:results_fdup}
\end{figure}

From FDup automatic clustering, we get 2,230 clusters, whose composition is reported in Figure~\ref{fig:results_fdup}.
Most of them (90.7\%) involves two profiles. This time, inward duplicate profiles (i.e., within the same registry) are immediately visible: 5.4\% of the clusters are composed of duplicates within ROAR, and 0.7\% of duplicates within OpenDOAR. Also, one duplicate within re3data is present. 
As before, most of the remaining clusters (69.04\%) are composed of ROAR and OpenDOAR profiles, then 21.02\% of FAIRsharing and re3data profiles, and the last 3.3\% involve a re3data profile with another one from OpenDOAR or ROAR. 
The 8.34\% of the clusters comprise three profiles, 69.89\% of which contain more than one profile from the same registry. 
Eight clusters are composed entirely of profiles within ROAR, which, together with OpenDOAR, is the registry with the highest number of multiple internal registrations for the same repository. 
The remaining 30.11\% is composed of a combination of three registries out of four.
The last 0.96\% of the clusters are composed of four or five profiles, six of which have at least one profile for each repository, and only five have no duplicate from the same registry. 

The exploitation of clusters to extend the duplicate sets produced 208 unique sets\footnote{The extended duplicate sets are 239, but 31 have been extended via merging: more than one set match with the same cluster, e.g., (\rr{978}, \rr{976}, \rr{5221}, \rr{2328}, \od{239}, \od{241}) via (\rr{976}, \rr{978}) from de-duplication, and (\od{241}, \rr{978})  and (\od{239}, \rr{2328}, \rr{5221}, \rr{976}) from registries, thus producing one single set}
(e.g., (\od{4194}, \rd{r3d100011201}, \fs{2560}) extending (\fs{2560}, \rd{r3d100011201})), 428 duplicate sets are obtained via de-duplication only, and 1,400 clusters overlap completely with the corresponding duplicate set, while 1,720 duplicate sets come from registry claims only.
The 70.7\% of the FDup-extended sets are obtained by adding one profile, 21.7\% by adding two profiles, and the remaining by adding up to five profiles, getting two sets of cardinality 8. 

Please note that the extended sets may not be complete. As an example, consider (\od{2373}, \rr{2115}, \rr{3755}, \rr{3591}, \rr{4562}) and  (\od{2373}, \rr{3591}, \rr{4562}, \rr{4695}) which are completely contained in the first set but \rr{4695}. 
The first one is obtained via the set (\rr{3591}, \rr{4562}, \od{2373}) and the cluster (\rr{4695}, \rr{3591}) and the second via the sets (\rr{3591}, \rr{4562}, \od{2373}) and (\rr{3755}, \od{2115}), and the cluster (\rr{4562}, \od{2115}). This was because we had two different clusters for the same repository, which were registered with different names. 
 
Assuming that the claims from the registries are correct, we have manually validated all the duplicate sets coming from FDup only or extended via a cluster to verify their correctness. 
Those extended via registry claims are also checked to understand why the FDup could not cluster them correctly.
To check the correctness of a given set, we considered the repository URLs and names as they were in our initial data.
When both are the same, the repository is considered the same. When one of the two is considerably different, we inspected the URLs and registry content as a tie-break. 
Incidentally, some repositories are no longer in the registry (7 among those inspected in the repository). These repositories come from claims among registries (ROAR to OpenDOAR mainly). There are also 16 wrong duplicate sets (9 because of FDup being wrong, 7 because of incorrect claims in the registries), and 74 not working URLs.
Furthermore, different duplicate sets can still refer to the same repository: de-duplication and registry claims do not have a common profile on whose basis our methodology can extend the duplicate set. Although we experimentally verified the presence of four of them, we cannot provide their exact number.

\section {Discussion}
On the one hand, the registry claims are a good starting point to identify multiple registrations of the same repository, but they are far from complete, and, most importantly, they are sometimes proved to be inaccurate.
On the other hand, a systematic procedure to automatically extend this claim set can be of aid, but can lead to incomplete duplicate sets or even disjoint and wrong ones.

The best of the two worlds can be achieved by combining the two approaches, but human intervention is still needed to obtain duplicate sets as accurately as possible; however, manual inspection is expensive and entails other known problems. 
Moreover, in some cases, time and effort do not suffice, as domain expert knowledge is required to determine whether two profiles can be considered the same (e.g., \rd{r3d100010553} and \fs{1956} whose website URLs resolve to very different pages). 
Furthermore, some choices would also remain discretionary, for example, in the case of two repositories being registered with the same name and with URLs resolving, in one case to a list of collections or subsets of a repository, and in the other to one of these subsets.
Unfortunately, there is no clear way to address such a conundrum without domain expert knowledge.

In conclusion, this study and the results presented here outline a first-of-its-kind wake-up call to raise awareness about the inherent ambiguity residing in scholarly repository registries.
While information scattering, consistency and duplication are not new in data quality literature, siloed, out-of-sync, scholarly repository registries can have a detrimental impact on the global scientific track record.

Firstly and foremost, inconsistency and incompleteness of a registry content can negatively affect its image and directly hamper its uptake and reliability in the eyes of the research community(ies) of reference.

Moreover, scholarly registries' downstream users, such as researchers, scholarly service providers, and Open Science infrastructures listing and aggregating their content, are exposed to potentially conflictual information, which has to be reconciled on an arbitrary (possibly manual) basis.
The OpenAIRE infrastructure, for example, aggregates data sources (i.e., repositories) from a variety of registries, such as FAIRsharing, OpenDOAR, re3data, and CRIS Systems. When the same repository is ingested from multiple registries, OpenAIRE needs to reconcile the duplicates into one single record. One copy is elected as master (if possible) and is enriched with all the identifiers from the other copies. As it is not always possible to choose a master, multiple copies of the same data source can still appear in the OpenAIRE portal. The de-duplication of data sources within OpenAIRE mostly entails manual work, and the cross-references provided by the registries and the de-duplication have to be checked and enriched with other repositories that were not included to get duplicate sets as accurate as possible. 

Finally, as subsequent registrations yield multiple identifiers (e.g., a DOI) for the same repository, assessing metrics reflecting the usage, adoption rate, and impact of a repository across the academic community can be hindered if equivalence across different registrations cannot be precisely pinpointed.

In our opinion, such problems can be solved mostly via an agreed-upon solution, paving the way towards the full support of interoperability across registries to enable a seamless exchange of the wealth of information contained therein.

\begin{acknowledgments}
This work was partially funded by the EC H2020 OpenAIRE-Nexus (Grant agreement 101017452).
\end{acknowledgments}
%
%
%
\bibliography{biblio}

\begin{thebibliography}{14}
\expandafter\ifx\csname natexlab\endcsname\relax\def\natexlab#1{#1}\fi
\providecommand{\url}[1]{\texttt{#1}}
\providecommand{\href}[2]{#2}
\providecommand{\path}[1]{#1}
\providecommand{\DOIprefix}{doi:}
\providecommand{\ArXivprefix}{arXiv:}
\providecommand{\URLprefix}{URL: }
\providecommand{\Pubmedprefix}{pmid:}
\providecommand{\doi}[1]{\href{http://dx.doi.org/#1}{\path{#1}}}
\providecommand{\Pubmed}[1]{\href{pmid:#1}{\path{#1}}}
\providecommand{\bibinfo}[2]{#2}
\ifx\xfnm\relax \def\xfnm[#1]{\unskip,\space#1}\fi
\bibitem[{Pampel et~al.(2013)Pampel, Vierkant, Scholze, Bertelmann, Kindling,
  Klump, Goebelbecker, Gundlach, Schirmbacher, and Dierolf}]{pampel2013}
\bibinfo{author}{H.~Pampel}, \bibinfo{author}{P.~Vierkant},
  \bibinfo{author}{F.~Scholze}, \bibinfo{author}{R.~Bertelmann},
  \bibinfo{author}{M.~Kindling}, \bibinfo{author}{J.~Klump},
  \bibinfo{author}{H.-J. Goebelbecker}, \bibinfo{author}{J.~Gundlach},
  \bibinfo{author}{P.~Schirmbacher}, \bibinfo{author}{U.~Dierolf},
\newblock \bibinfo{title}{Making {{Research Data Repositories Visible}}:
  {{The}} re3data.org {{Registry}}},
\newblock \bibinfo{journal}{PLOS ONE} \bibinfo{volume}{8}
  (\bibinfo{year}{2013}) \bibinfo{pages}{e78080}.
  \DOIprefix\doi{10.1371/journal.pone.0078080}.
\bibitem[{Wallis et~al.(2013)Wallis, Rolando, and Borgman}]{wallis2013}
\bibinfo{author}{J.~C. Wallis}, \bibinfo{author}{E.~Rolando},
  \bibinfo{author}{C.~L. Borgman},
\newblock \bibinfo{title}{If {{We Share Data}}, {{Will Anyone Use Them}}? data
  {{Sharing}} and {{Reuse}} in the {{Long Tail}} of {{Science}} and
  {{Technology}}},
\newblock \bibinfo{journal}{PLoS ONE} \bibinfo{volume}{8}
  (\bibinfo{year}{2013}). \DOIprefix\doi{10.1371/journal.pone.0067332}.
\bibitem[{Davidson et~al.(2014)Davidson, Jones, and Molloy}]{davidson2014}
\bibinfo{author}{J.~Davidson}, \bibinfo{author}{S.~Jones},
  \bibinfo{author}{L.~Molloy},
\newblock \bibinfo{title}{Big data: The potential role of research data
  management and research data registries},
\newblock in: \bibinfo{booktitle}{{{IFLA WLIC}} 2014}, \bibinfo{address}{{Lyon,
  France}}, \bibinfo{year}{2014}.
\bibitem[{Pasquetto et~al.(2019)Pasquetto, Borgman, and
  Wofford}]{pasquetto2019}
\bibinfo{author}{I.~V. Pasquetto}, \bibinfo{author}{C.~L. Borgman},
  \bibinfo{author}{M.~F. Wofford},
\newblock \bibinfo{title}{Uses and {{Reuses}} of {{Scientific Data}}: The
  {{Data Creators}}' {{Advantage}}},
\newblock \bibinfo{journal}{Harvard Data Science Review} \bibinfo{volume}{1}
  (\bibinfo{year}{2019}). \DOIprefix\doi{10.1162/99608f92.fc14bf2d}.
\bibitem[{Silvello(2018)}]{silvello2018}
\bibinfo{author}{G.~Silvello},
\newblock \bibinfo{title}{Theory and {{Practice}} of {{Data Citation}}},
\newblock \bibinfo{journal}{Journal of the Association for Information Science
  and Technology} \bibinfo{volume}{69} (\bibinfo{year}{2018})
  \bibinfo{pages}{6--20}. \DOIprefix\doi{10.1002/asi.23917}.
  \href{http://arxiv.org/abs/1706.07976}{{\tt arXiv:1706.07976}}.
\bibitem[{Ball et~al.(2014)Ball, Ashley, McCann, Molloy, and Van
  Den~Eynden}]{ball2014}
\bibinfo{author}{A.~Ball}, \bibinfo{author}{K.~Ashley},
  \bibinfo{author}{P.~McCann}, \bibinfo{author}{L.~Molloy},
  \bibinfo{author}{V.~Van Den~Eynden},
\newblock \bibinfo{title}{Show me the data: The pilot {{UK Research Data
  Registry}}},
\newblock \bibinfo{journal}{International Journal of Digital Curation}
  \bibinfo{volume}{9} (\bibinfo{year}{2014}) \bibinfo{pages}{132--141}.
  \DOIprefix\doi{10.2218/ijdc.v9i1.307}.
\bibitem[{{Mart{\'i}n-Mart{\'i}n} et~al.(2020){Mart{\'i}n-Mart{\'i}n},
  Thelwall, {Orduna-Malea}, and {Delgado
  L{\'o}pez-C{\'o}zar}}]{martin-martin2020}
\bibinfo{author}{A.~{Mart{\'i}n-Mart{\'i}n}}, \bibinfo{author}{M.~Thelwall},
  \bibinfo{author}{E.~{Orduna-Malea}}, \bibinfo{author}{E.~{Delgado
  L{\'o}pez-C{\'o}zar}},
\newblock \bibinfo{title}{Google {{Scholar}}, {{Microsoft Academic}},
  {{Scopus}}, {{Dimensions}}, {{Web}} of {{Science}}, and {{OpenCitations}}'
  {{COCI}}: A multidisciplinary comparison of coverage via citations},
\newblock \bibinfo{journal}{Scientometrics}  (\bibinfo{year}{2020}).
  \DOIprefix\doi{10.1007/s11192-020-03690-4}.
  \href{http://arxiv.org/abs/2004.14329}{{\tt arXiv:2004.14329}}.
\bibitem[{Gusenbauer and Haddaway(2020)}]{gusenbauer2020}
\bibinfo{author}{M.~Gusenbauer}, \bibinfo{author}{N.~R. Haddaway},
\newblock \bibinfo{title}{Which academic search systems are suitable for
  systematic reviews or meta-analyses? {{Evaluating}} retrieval qualities of
  {{Google Scholar}}, {{PubMed}}, and 26 other resources},
\newblock \bibinfo{journal}{Research Synthesis Methods} \bibinfo{volume}{11}
  (\bibinfo{year}{2020}) \bibinfo{pages}{181--217}.
  \DOIprefix\doi{10.1002/jrsm.1378}.
\bibitem[{Visser et~al.(2021)Visser, {van Eck}, and Waltman}]{visser2021}
\bibinfo{author}{M.~Visser}, \bibinfo{author}{N.~J. {van Eck}},
  \bibinfo{author}{L.~Waltman},
\newblock \bibinfo{title}{Large-scale comparison of bibliographic data sources:
  {{Scopus}}, {{Web}} of {{Science}}, {{Dimensions}}, {{Crossref}}, and
  {{Microsoft Academic}}}  (\bibinfo{year}{2021}) \bibinfo{pages}{22}.
  \DOIprefix\doi{10.1162/qss\_a\_00112}.
\bibitem[{Harzing(2019)}]{harzing2019}
\bibinfo{author}{A.~W. Harzing},
\newblock \bibinfo{title}{Two new kids on the block: {{How}} do {{Crossref}}
  and {{Dimensions}} compare with {{Google Scholar}}, {{Microsoft Academic}},
  {{Scopus}} and the {{Web}} of {{Science}}?},
\newblock \bibinfo{journal}{Scientometrics} \bibinfo{volume}{120}
  (\bibinfo{year}{2019}) \bibinfo{pages}{341--349}.
  \DOIprefix\doi{10.1007/s11192-019-03114-y}.
\bibitem[{Aryani et~al.(2020)Aryani, Fenner, Manghi, Mannocci, and
  Stocker}]{aryani2020}
\bibinfo{author}{A.~Aryani}, \bibinfo{author}{M.~Fenner},
  \bibinfo{author}{P.~Manghi}, \bibinfo{author}{A.~Mannocci},
  \bibinfo{author}{M.~Stocker},
\newblock \bibinfo{title}{Open {{Science Graphs Must Interoperate}}!},
\newblock in: \bibinfo{booktitle}{{{ADBIS}}, {{TPDL}} and {{EDA}} 2020 {{Common
  Workshops}} and {{Doctoral Consortium}}}, \bibinfo{publisher}{{Springer}},
  \bibinfo{year}{2020}.
\bibitem[{Sansone et~al.(2019)Sansone, McQuilton, {Rocca-Serra},
  {Gonzalez-Beltran}, Izzo, Lister, and Thurston}]{sansone2019}
\bibinfo{author}{S.-A. Sansone}, \bibinfo{author}{P.~McQuilton},
  \bibinfo{author}{P.~{Rocca-Serra}}, \bibinfo{author}{A.~{Gonzalez-Beltran}},
  \bibinfo{author}{M.~Izzo}, \bibinfo{author}{A.~L. Lister},
  \bibinfo{author}{M.~Thurston},
\newblock \bibinfo{title}{{{FAIRsharing}} as a community approach to standards,
  repositories and policies},
\newblock \bibinfo{journal}{Nature Biotechnology} \bibinfo{volume}{37}
  (\bibinfo{year}{2019}) \bibinfo{pages}{358--367}.
  \DOIprefix\doi{10.1038/s41587-019-0080-8}.
\bibitem[{De~Bonis et~al.(2022)De~Bonis, Manghi, and Atzori}]{de2022fdup}
\bibinfo{author}{M.~De~Bonis}, \bibinfo{author}{P.~Manghi},
  \bibinfo{author}{C.~Atzori},
\newblock \bibinfo{title}{Fdup: a framework for general-purpose and efficient
  entity deduplication of record collections},
\newblock \bibinfo{journal}{PeerJ Computer Science} \bibinfo{volume}{8}
  (\bibinfo{year}{2022}) \bibinfo{pages}{e1058}.
\bibitem[{Manghi et~al.(2021)Manghi, Atzori, Bardi, Baglioni, Schirrwagen,
  Dimitropoulos, La~Bruzzo, Foufoulas, Mannocci, Horst, Czerniak, Kiatropoulou,
  Kokogiannaki, De~Bonis, Artini, Ottonello, Lempesis, Ioannidis, Manola, and
  Principe}]{manghi2021}
\bibinfo{author}{P.~Manghi}, \bibinfo{author}{C.~Atzori},
  \bibinfo{author}{A.~Bardi}, \bibinfo{author}{M.~Baglioni},
  \bibinfo{author}{J.~Schirrwagen}, \bibinfo{author}{H.~Dimitropoulos},
  \bibinfo{author}{S.~La~Bruzzo}, \bibinfo{author}{I.~Foufoulas},
  \bibinfo{author}{A.~Mannocci}, \bibinfo{author}{M.~Horst},
  \bibinfo{author}{A.~Czerniak}, \bibinfo{author}{K.~Kiatropoulou},
  \bibinfo{author}{A.~Kokogiannaki}, \bibinfo{author}{M.~De~Bonis},
  \bibinfo{author}{M.~Artini}, \bibinfo{author}{E.~Ottonello},
  \bibinfo{author}{A.~Lempesis}, \bibinfo{author}{A.~Ioannidis},
  \bibinfo{author}{N.~Manola}, \bibinfo{author}{P.~Principe},
  \bibinfo{title}{{{OpenAIRE Research Graph Dump}}}, \bibinfo{year}{2021}.
  \DOIprefix\doi{10.5281/zenodo.5801283}.

\end{thebibliography}

\end{document}